\documentclass[%
 reprint,
 amsmath,amssymb,
 aps,
]{revtex4-1}

\usepackage{graphicx}
\usepackage{dcolumn}
\usepackage{bm}

\newcommand{\Lb}{\left(}
\newcommand{\Rb}{\right)}
\newcommand{\pom}{I\!\!P}
\begin{document}


\title{ Amplitudes and Cross-sections at the LHC }%

\author{Errol Gotsman}
\affiliation{School of Physics and Astronomy, Tel Aviv University, Tel 
Aviv, Israel}


\date{\today}

\begin{abstract}
We describe  the elements of the GLM model that successfully
describes soft
hadronic interactions at energies from ISR to LHC. This model is based on
a
single Pomeron with a large intercept $\Delta_{\pom}$ = 0.23 and slope
$\alpha_{\pom}'$ = 0, and so provides a natural matching with perturbative
QCD. We analyze the elastic, single diffractive and double diffractive 
amplitudes, and compare the behaviour of the GLM amplitudes to those of 
other parameterizations.
 We summarize the main features and results of  competing
models for soft interactions  at LHC energies.
\end{abstract}

\maketitle
\subsection{Introduction}
   The recent measurements  of the proton-proton cross sections    at the
LHC at an energy of $W$ = 7 TeV, allows one to appraise
the numerous models that have been proposed to describe soft interactions.
  The classical Regge pole model {\it \`a la} Donnachie and Landshoff
\cite{DL1}, which provided
 a reasonable description of soft hadron-hadron scattering upto
the Tevatron energy,  fails when extended to LHC energies
\cite{DL2}.
In addition it has the intrinsic problem of violating the
Froissart-Martin bound   
\cite{FM}.

  At present there are a number of models based on Reggeon Field
Theory that provide an acceptable
 description of proton-proton scattering data over the energy
range from ISR to LHC. I will describe the essential features of the GLM
model \cite{GLM1} as an example of a model of this type, before comparing
its results with other competing models on the market.
 \subsection{Basic features of the GLM model}
We utilize the simple    
two channel Good-Walker (GW) \cite{GW} model,
        to account for elastic scattering and for diffractive
dissociation into states with masses
that are much smaller than the initial energy.
and impose the unitarity constraint
  by requirying that
$$
2\,\mbox{Im}\,A_{i,k}\left(s,b\right)=|A_{i,k}\left(s,b\right)|^2
+G^{in}_{i,k}(s,b)
$$
where, $A_{i,k}$ denotes the diagonalized interaction amplitude and
$G^{in}_{i,k}$,  the contribution of all
non GW inelastic processes.\\
 A general solution for the amplitude satisfying the above unitarity
equation is:
\begin{equation}
A_{i,k}(s,b)=i \Lb 1- \exp\Lb - \frac{\Omega_{i,k}(s,b)}{2}\Rb\Rb
\end{equation}
 the opacities $\Omega_{i,k}$  are arbitrary.
In the eikonal approximation $\Omega_{i,k}$
are assumed to be  real, and taken to be the
contribution of a single Pomeron exchange.

GLM parameterize the opacity : $$ \Omega_{i,k}(s,b)
\,\,=\,\,
g_i(b)\,g_k(b)\,P(s)$$ where
$P(s)\,=\,s^\Delta$, and $g_i(b)$ and $g_k(b)$ are the
Pomeron-hadron vertices given by:
$$ g_l\Lb
b\Rb\,=\,g_l\,S_l(b)\,=\,\frac{g_l}{4 \pi}\,m^3_l \,b\,K_1\Lb m_l b\Rb.
$$
$S_l(b)$ is the Fourier transform of $\frac{1}{(1 + q^2/m^2_l)^2}$, \\
where $q$ is the transverse momentum carried by the Pomeron,
$l=i,k$.
 The form of $P(s)$ used by GLM, corresponds to
a Pomeron trajectory slope $\alpha'_{\pom}$ = 0.
This is compatible with the exceedingly small fitted value
of $\alpha_{\pom}^{\prime}$, (0.028 GeV$^{-2}$) and in accord with $N$=4
SYM.
\par
For the case of $\Delta_{\pom} \to 0$,  the Pomeron
interaction
leads to a new source of diffraction production with large mass
($M \propto s$), which cannot be described by the Good-Walker mechanism.
Taking $\alpha'_{\pom}$ = 0 , allows one to sum all diagrams
having   Pomeron interactions~\cite{GLM11,GLM12}. This is the
advantage
of such an approach. The GLM model  only takes into account  triple
Pomeron
interaction vertices ($G_{3\pom}$), this provides a natural matching to 
the hard Pomeron, since  at short distances $G_{3\pom} \propto
\alpha_{s}^2$,
while other vertices are much smaller.
 A full description of the procedure for summing all diagrams
(enhanced + semi-enhanced) is contained in
 \cite{GLM11,GLM12,GLMLAST}.
We would like to emphasize that in  the GLM model, the GW sector
contributes to
both low and high
diffracted mass, while the non-GW sector contributes only to
high mass diffraction ($\log\Lb M^2/s_0\Rb \approx\,1/\Delta_{\pom}$).
\par The GLM model has 14 parameters describing the Pomeron and Reggeon
sectors. The values of these parameters are determined by   fitting to
data for $\sigma_{tot}$, $\sigma_{el}$, $\sigma_{sd}$, $\sigma_{dd}$ and
$B_{el}$ in the ISR-LHC range \cite{GLMLAST}. We find the best fit value
for
$\alpha_{\pom}$ = 0.21, however to be in accord with the LHC data we have
tuned $\alpha_{\pom}$ to 0.23. The fitted values for $\alpha'_{\pom}$
is 0.028 GeV$^{-2}$, while the triple Pomeron vertex $G_{3\pom}$ = 0.03
GeV$^{-1}$.
\subsection{Experimental Data and GLM results}
The comparison of our results with experimental data
$\sigma_{tot}$, $\sigma_{el}$ and for $B_{el}$ is shown in Fig.~1, 2 and 
3.
The results for $\sigma_{inel}$, $\sigma_{sd}$ and $\sigma_{dd}$, are 
given 
in Fig.4 which is taken from 
 the talk given by Orlando Villalobos Baillie (for the Alice
collaboration) (see reference \cite{Villa4}), where
 the experimental data,  our results and the results of other models 
  are displayed.

To summarize our results at high energy, we obtain an
excellent
reproduction of TOTEM's values for $\sigma_{tot}$ and $\sigma_{el}$.
The quality of our good fit to $B_{el}$ is maintained.
As regards $\sigma_{inel}$, our results are in accord with the higher
values obtained by ALICE~\cite{ALICE} and TOTEM~\cite{TOTEM};
 ATLAS~\cite{ATLAS} and CMS~\cite{CMS} quote lower values with
large extrapolation errors, see~\cite{JPR}.
 We refer the reader to
\cite{JPR} who suggests
that the lower values found by ATLAS and CMS maybe due to the simplified
Monte Carlo that they used to estimate their diffractive background.
\par
There are also recent results at $W$ = 57 TeV by the Auger
Collaboration~\cite{auger} for $\sigma_{tot}$ and $\sigma_{inel}$.   
See Table I for a comparison of experimental results at W = 7 and 57 TeV 
and the GLM model.
\begin{figure}
\begin{flushleft}
\includegraphics[width=0.3\textwidth]{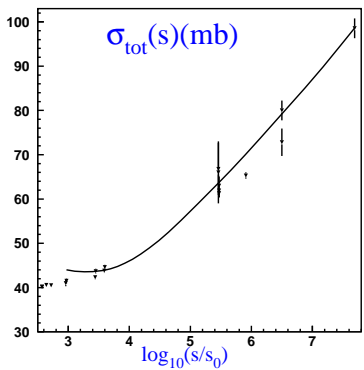} 
\caption{\label{fig:1} The GLM results compared to data for
$\sigma_{tot}$.}
\end{flushleft}
\end{figure}

\begin{figure}
\begin{flushright}
\includegraphics[width=0.3\textwidth]{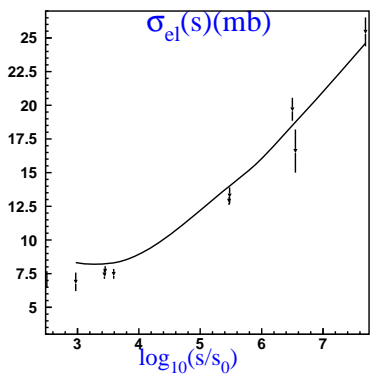}
\caption{\label{fig:2} The GLM results compared to data for
$\sigma_{el}$.}
\end{flushright} 
\end{figure}

\begin{figure}
\begin{flushleft}
\includegraphics[width=0.3\textwidth]{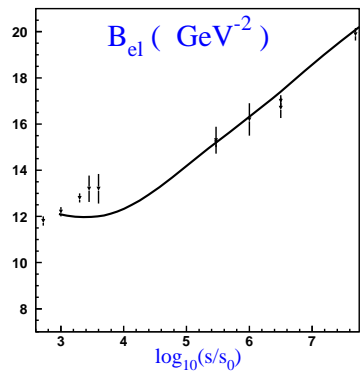}
\caption{\label{fig:3} The GLM results compared to data for
 $B_{el}$.}
\end{flushleft}
\end{figure}

\begin{table*}
\begin{tabular}{|l |l|l|l|l|}
\hline \hline
W & $\sigma^{model}_{tot}$(mb) &  $\sigma^{exp}_{tot}$(mb) &
$\sigma^{model}_{el}$(mb)&
 $\sigma^{exp}_{el}$(mb)\\
\hline
7 TeV~~~ & 98.6 ~~~~& TOTEM: 98.6 $\pm$2.2
~~~~ & 24.6 ~~~~~ & TOTEM: 25.4$\pm 1.1$
 ~~~~~~\\
\hline
\end{tabular}

\begin{tabular}{|l|l|l|l|l|}
\hline
 W & $\sigma^{model}_{in}(mb)$&
$\sigma^{exp}_{in}$(mb) &$B^{model}_{el}(GeV^{-2}) $&
$B^{exp}_{el}(GeV^{-2})$\\
\hline
7 TeV & 74.0  & CMS: 68.0$\pm2^{syst}\pm 2.4^{lumi}\pm 4^{extrap}$ &
 20.2 &
TOTEM: 19.9$  \pm 0.3\,$\\
 &  &  ATLAS: 69.4$\pm 2.4^{exp}  \pm 6.9^{extrap}$  & &\\
& &  ALICE: 73.2 $(+2./-4.6)^{model}  \pm 2.6^{lumi}$  & & \\
& & TOTEM: 73.5  $\pm 0.6^{stat}  \pm 1.8^{syst}$ & & \\
\hline
\end{tabular}

\begin{tabular}{|l |l|l|l|l|}
\hline
W & $\sigma^{model}_{sd}(mb)$ &  $\sigma^{exp}_{sd}$(mb) &
 $\sigma^{model}_{dd}$(mb)& $\sigma^{exp}_{dd}$(mb)\\
\hline
7 TeV & 10.7${}^{GW}$ +  4.18${}^{nGW}$ ~ & ALICE : 14.9(+3.4/-5.9)
 & 6.21${}^{GW}$  + 1.24${}^{nGW}$  ~~~~~~~& ALICE: 9.0 $\pm$ 2.6 ~\
\\
\hline\hline 
\end{tabular}

\begin{tabular}{|l |l|l|}
\hline \hline
W & $\sigma^{model}_{tot}$(mb) &  $\sigma^{exp}_{tot}$(mb) \\ \hline
57 TeV~~~ & 130 ~~~~& AUGER: 133
$\pm 13^{stat} \pm 17^{sys} \pm 16^{Glauber}$  \\ \hline\hline
& $\sigma^{model}_{inel}$(mb)& $\sigma^{exp}_{inel}$(mb)\\
\hline
~~~ & 95.2 ~~~~~ & AUGER: 92 $\pm7^{stat}\pm11^{syst} \pm7^{Glauber}$
 ~~~~~~\\
\hline\hline
\end{tabular}
{\caption. ~Comparison of the values obtained from the GLM model with 
experimental results at W = 7 and 57 TeV.}
\end{table*}

\subsection{Alternative Models}
There are several models on the market today that manage to reproduce the
LHC experimental results. The most promising of these are summarized here,
and their results are compared with those of GLM~\cite{GLM1} in Table~I.
\par
The Durham group's approach for describing soft hadron-hadron 
scattering~\cite{RMK1} is similar  to the GLM~\cite{GLM1} approach, they
include both
enhanced and semi-enhanced diagrams. The two groups utilize different
techniques for summing the multi-Pomeron diagrams. The Durham group have a
bare (prior to screening) QCD Pomeron, with intercept $\Delta_{bare}$ =
0.32. This model~\cite{RMK1} which was tuned to describe collider data,
predicts values for $\sigma_{tot}$, $\sigma_{el}$ and $\sigma_{inel}$, 
which
are lower than the TOTEM~\cite{TOTEM} data.
 To be consistent with the TOTEM results, RMK~\cite{RMK2}  have proposed
an alternative formulation, based on
a 3-channel eikonal description, with 3 diffractive eigenstates of
different sizes, but with only one Pomeron whose intercept and slope are:
$\Delta_{\pom}$ = 0.14; $\alpha^{'}_{\pom}$ = 0.1 GeV$^{-2}$. Their  
results are
shown in Table~II in the column KMR2.
\par
Ostapchenko~\cite{OS} [pre LHC] has made a comprehensive
calculation
 in the framework of Reggeon Field Theory,
 based on the resummation
 of both enhanced and semi-enhanced Pomeron diagrams.
To fit the total and diffractive cross sections he assumes two Pomerons: 
(for his solution set C)
"Soft Pomeron"
 $\,\,\,\alpha^{Soft} = 1.14\, +\, 0.14t$ and a
"Hard Pomeron"
 $\,\,\,\alpha^{Hard} = 1.31\, +\, 0.085t$. His results are
quoted in Table~II, in the column Ostap(C).
\par
Kaidalov-Poghosyan~\cite{KP}
have a model which is based on Reggeon calculus,
they attempt to describe data on soft diffraction taking into account all
possible non-enhanced absorptive corrections to 3 Reggeon vertices and
loop diagrams.
  It is a single Pomeron  model and with secondary  Regge
poles, their Pomeron has the following intercept and slope:
 $\Delta_{\pom}\;=\;0.12$ and
$\alpha^{'}_{\pom}\;=\;0.22$ GeV$^{-2}$. Their results are shown in 
Table~II, in the column KP.
\par
Ciesielski and Goulianos have proposed an "event generator"~\cite{BMR} 
which    
 is based on the BMR-enhanced PYTHIA8 simulation. In
Table~II their results are denoted by BMR.

\begin{figure*}
\includegraphics[width=0.99\textwidth]{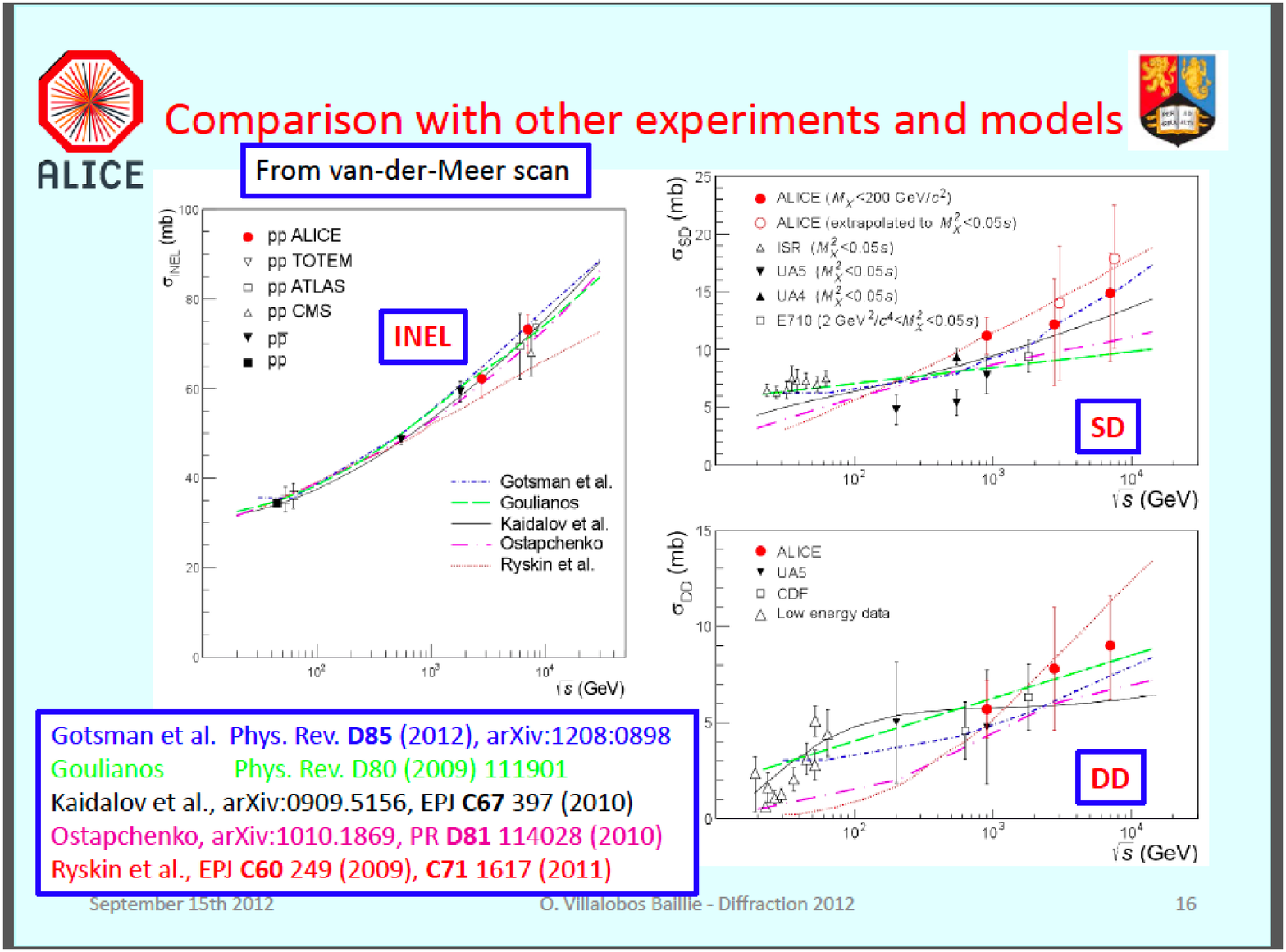}
\caption{\label{fig:Villa} Comparison of Models with LHC data from
Villalabos Ballie's talk at Diffraction 2012. \cite{Villa4}}
\end{figure*}

\begin{table}[tb]
\begin{tabular}{|c|c|c|c|c|c|c|}
\hline
 W = 1.8 TeV  & GLM & KMR2 & Ostap(C)  & BMR$^{*}$ 
&
KP
\\
\hline
$\sigma_{\rm tot}(mb)$ & 79.2 & 79.3 & 73.0 & 81.03 & 75.0
\\ \hline
$\sigma_{\rm el}(mb)$ & 18.5 & 17.9 & 16.8 & 19.97 & 16.5   \\
\hline
$\sigma_{SD}(mb)$ & 11.27 & 5.9(LM) & 9.2 & 10.22 &10.1   \\ \hline
$\sigma_{DD}(mb)$ & 5.51 & 0.7(LM) & 5.2 & 7.67 & 5.8     \\ \hline
$B_{el}(GeV^{-2})$ & 17.4 & 18.0 & 17.8 &  &    \\
\hline
\end{tabular}

\begin{tabular}{|c|c|c|c|c|c|c|}
\hline

 $W$ = 7 TeV  & GLM & KMR2 & Ostap(C)  & BMR & KP
\\
\hline
$\sigma_{\rm tot}$(mb) & 98.6 & 97.4 & 93.3 & 98.3 & 96.4
\\ \hline  
$\sigma_{\rm el}$(mb) & 24.6 & 23.8 & 23.6 & 27.2 & 24.8    \\
\hline
$\sigma_{SD}$(mb) & 14.88 & 7.3(LM) & 10.3 & 10.91 &12.9   \\ \hline
$\sigma_{DD}$(mb) & 7.45 & 0.9(LM) & 6.5 & 8.82 & 6.1    \\ \hline
$B_{el}$(GeV$^{-2})$ & 20.2 & 20.3 & 19.0 &  &19.0    \\
                                                                         
\hline
\end{tabular}

\begin{tabular}{|c|c|c|c|c|c|c|}
\hline
 $W$ = 14 TeV  & GLM & KMR2 & Ostap(C)  & BMR & KP   \\ 
\hline
$\sigma_{tot}$(mb) & 109.0 & 107.5 & 105. & 109.5 & 108. \\ \hline
$\sigma_{el}$(mb) & 27.9 & 27.2 & 28.2 & 32.1 & 29.5  \\ \hline
$\sigma_{SD}$(mb) & 17.41 & 8.1(LM) & 11.0 & 11.26 & 14.3   \\ \hline
$\sigma_{DD}$(mb) & 8.38 & 1.1(LM) & 7.1 & 9.47 & 6.4    \\ \hline
$B_{el}$(GeV$^{-2})$ & 21.6 & 21.6 & 21.4 &  & 20.5    \\
\hline
\end{tabular}
\label{tab:a}
\caption{Comparison of results of the different models for $W$ = 1.8,  7
and 14 TeV.}

\end{table}

\subsection{Amplitudes}

  Until recently  most of the comparison of models has been done 
on the level of cross-sections (which are areas), and only reveal the 
energy dependence, and  therefore are not very helpful to discriminate 
between the different models. Having the behaviour of the various 
amplitudes as functions of impact parameter (momentum transfer) would be 
more revealing.
  Unfortunately, there is a paucity of material available on amplitudes,
and most refer only to the elastic amplitude.

 In Fig.5. we show elastic amplitudes emanating from the 
GLM model for various energies. We note the overall gaussian shape of the 
elastic amplitudes for all energies 0.545 $\leq\; W \; \leq$ 57 TeV,
with the width and height of the gaussian growing with increasing energy. 
For 
small values of b the slope of the amplitudes decreases with increasing 
energy. The elastic amplitude (as $ b \rightarrow $ 0) becomes almost flat 
for W = 57 TeV, where it  is still below  
 the Unitarity limit $ A_{el} = 1$ .

In Fig.6 we show the elastic, single 
diffraction and double 
diffraction amplitudes as functions of b for W = 7 TeV.
 Note the completely different shapes of the three amplitudes, the 
elastic amplitude $A_{el}(b)$ is gaussian in shape, while the single 
diffractive amplitude $A_{sd}(b)$ and the double diffractive amplitude 
$A_{dd}(b)$ are very small at small b . $A_{sd}$ has a peak at 1.25 fm, 
while $A_{dd}$'s maximum is at b = 2.15 fm.

 \begin{figure}
\begin{flushleft}
\includegraphics[width=0.4\textwidth]{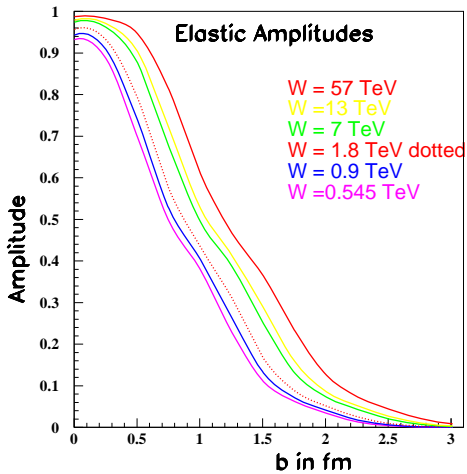}
\caption{\label{fig:5} The GLM elastic amplitudes for LHC energies. }
\end{flushleft}
\end{figure}  

\begin{figure}
\begin{flushright}
\includegraphics[width=0.4\textwidth]{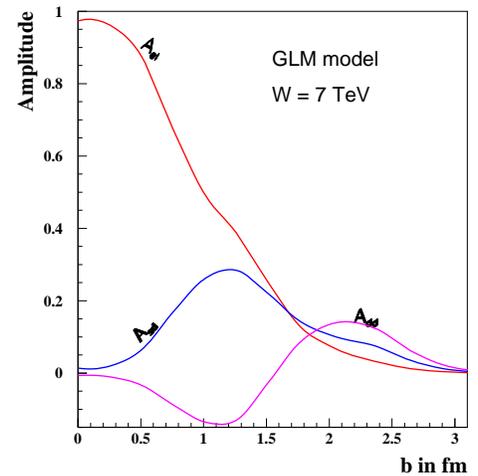}
\caption{\label{fig:6} The GLM elastic, single diffractive and double 
diffractive amplitudes for W = 7 TeV. }
\end{flushright}
\end{figure}

  The Durham group \cite{RMK2} have attempted to extract the form of the 
Elastic Opacity directly from the data.They assume that at high energies 
the 
real part of the scattering amplitude is very much smaller than the 
imaginary part, then to a good approximation
$$ A(b) = i [1 - \exp(-\Omega(b)/2)] $$
(see Eqn(1)).
 As $\Omega_{el} = - 2 ln(1 - A_{el})$, they determine the 
Opacity directly from the data since
$$
{\rm Im}A(b)~=~\int \sqrt{\frac{d\sigma_{\rm
el}}{dt}\frac{16\pi}{1+\rho^2}}~ J_0(q_tb)~ \frac{q_tdq_t}{4\pi},
$$
where $q_t=\sqrt{|t|}$ and $\rho \equiv {\rm Re}A/{\rm Im}A$.
Their results are shown in  Fig. 7.
  
\begin{figure}
\begin{flushleft}
\includegraphics[width=0.4\textwidth]{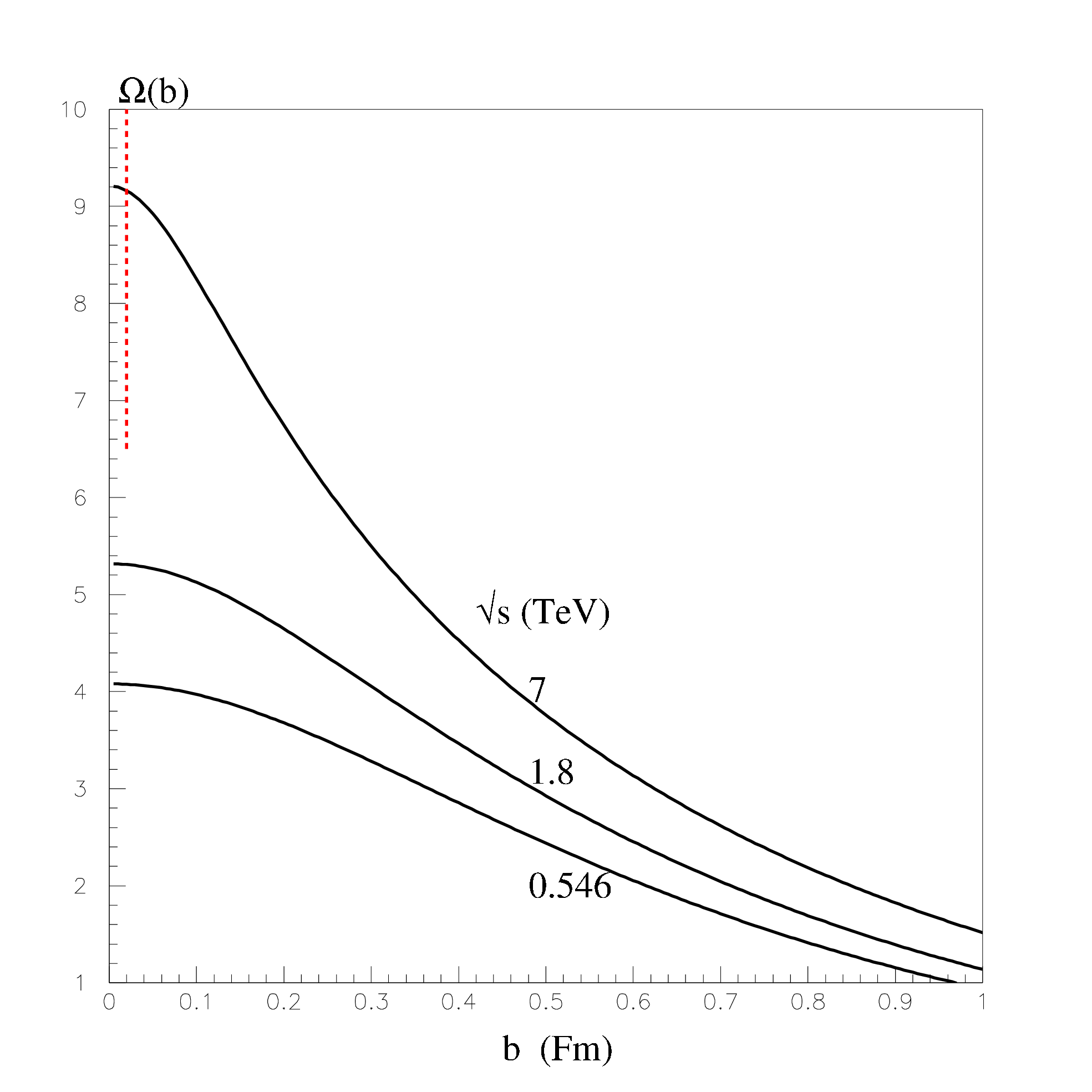}
\caption{\label{fig:7}  
The proton opacity $\Omega(b)$ determined directly from the
$pp$ $d\sigma_{\rm el}/dt$ data at 546 GeV , 1.8 TeV
and 7 TeV  data.  The uncertainty on the LHC value at $b=0$
is indicated by a dashed  red  line. This figure is taken from 
\cite{RMK2}} which should be consulted for details.
\end{flushleft} 
\end{figure}

\begin{figure}
\begin{flushright}
\includegraphics[width=0.4\textwidth]{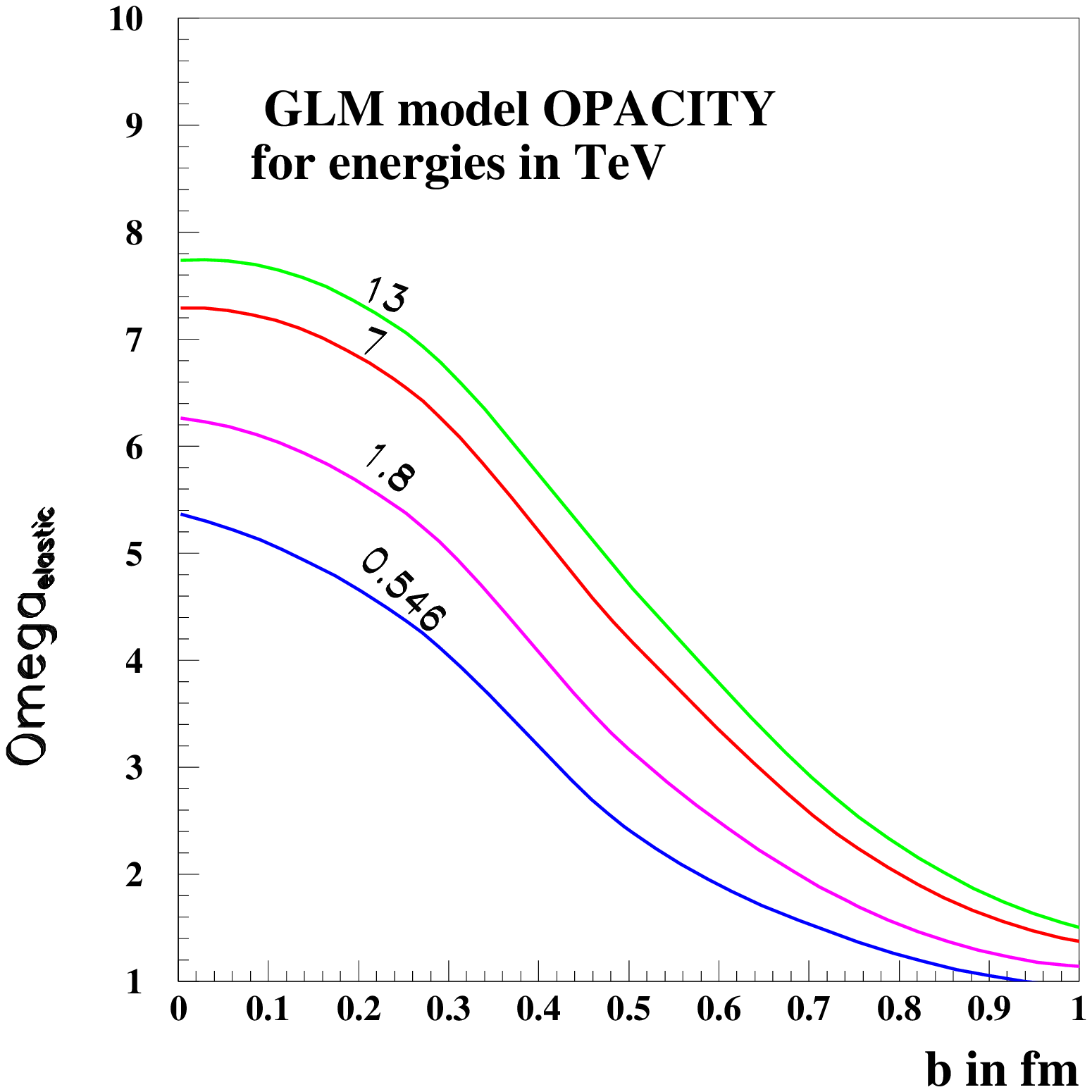}
\caption{\label{fig:8} Opacites calculated using the GLM model.}
\end{flushright}
\end{figure}

\par
The Durham group \cite{RMK2}  find that at $Sp\bar{p}S$
and Tevatron energies the Opacity distributions have
 appoximately a Gaussian form.
 The analogous GLM model results are shown in  Fig. 8, are in agreement 
with \cite{RMK2} regarding the shape of $\Omega_{el}(b)$, and in addition 
suggest that this is also true for the LHC 
energies. GLM find that with increasing energy, the intercept of the 
Opacity at b =0 increases,
while 
 the slope  at small b decreases. 

\par
Ferreira, Kodama and Kohara \cite{FKK}  have recently made a detailed 
study of the 
proton-proton elastic amplitude for center of mass energy W = 7 TeV, based 
on Stochastic Vacuum Model (see \cite{FKK} for more details). 

In Fig.9 we show the GLM and FKK elastic amplitudes as a function of the
impact parameter. Although the shapes are similar, the FKK amplitude is 
lower. If we normalize the FKK amplitude to the GLM value at b 
= 0, we note that the amplitudes which are gaussian in shape,  have very 
similar behaviour as a function 
of the impact parameter. 
\begin{figure}
\includegraphics[width=0.45\textwidth]{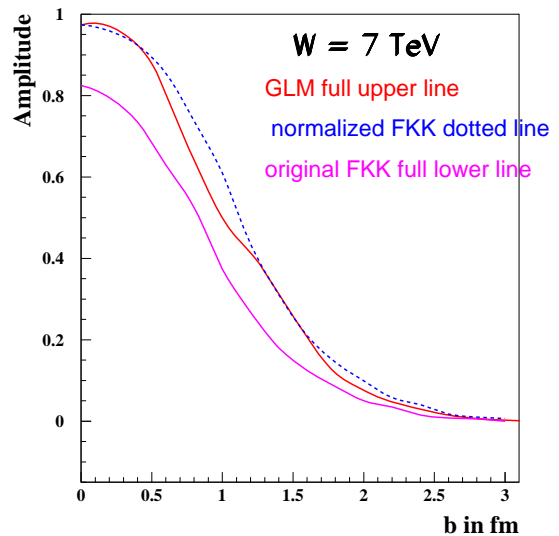}
\caption{\label{fig:9} Comparison of the elastic amplitude determined by 
FKK \cite{FKK} and the GLM model.}
\end{figure}

\subsection{Conclusions}

We~\cite{GLM1}
 have succeded in  building a model  for soft interactions,
which provides a very good description all high energy data, including the
LHC measurements.
 The model is based on a Pomeron with a
 large intercept ($\Delta_{\pom} = 0.23$) and very small slope
($\alpha'_{\pom}$ =0.028).
We find no need to introduce two Pomerons: i.e. a soft and a hard one.
The
Pomeron in our model provides a natural matching with the hard Pomeron
 in  processes that occur at short distances.
  The
 qualitative features of our model are close to what one expects
 from $N$=4 SYM~\cite{GLM11,GLM12}, which is the only
theory that is able to treat  long distance physics on a
 solid  theoretical basis.
\par
  In concluding
 I  appeal to all model builders and Monte Carlo advocates to
publish numerical values for their  amplitudes, as this would enable 
one to check the inherent differences between the various approaches to 
soft scattering.
 
\begin{acknowledgments}
 I would like to thank my colleagues and friends Evgeny  Levin and Uri 
Maor, for a fruitful collaboration.
\end{acknowledgments}

\end{document}